\documentclass[prl,showpacs,twocolumn,floatfix]{revtex4}
\usepackage{graphicx}

\def\bea{\begin{eqnarray}}
\def\eea{\end{eqnarray}}
\def\ben{\begin{equation}}
\def\een{\end{equation}}
\def\benu{\begin{enumerate}}
\def\enu{\end{enumerate}}

\def\n{n}

\def\sss{\scriptscriptstyle\rm}

\def\1var{(\bx_1...\bx\N)}

\def\br{{\bf r}}

\def\bx{{\br t}}

\def\x{_{\sss X}}
\def\c{_{\sss C}}
\def\s{_{\sss S}}
\def\xc{_{\sss XC}}

\def\N{_{\sss N}}

\def\LDA{^{\rm LDA}}

\def\PBE{^{\rm PBE}}

\def\sph_int{ {\int d^3 r}}

\input rotate

\begin{document}
\def\z{_\zeta}
\def\vec#1{{\bf #1}}

\title{Relevance of the slowly-varying electron gas to atoms, molecules, and solids}
\author{John P. Perdew}
\author{Lucian A. Constantin}
\author{Espen Sagvolden}
\affiliation{Department of Physics and Quantum Theory Group, Tulane University,
LA 70118}
\author{Kieron Burke}
\affiliation{Department of Chemistry \& Chemical Biology, Rutgers University, 610 Taylor Rd, Piscataway, NJ 08854}
\date{\today}
\begin{abstract} 
Electron densities become both large and slowly-varying in a certain
semi-classical limit: The local density approximation becomes
exact for both the kinetic and exchange energies,
while the leading
correction is given by the second-order gradient expansion.  But real
matter has nuclear cusps, which produce ``Scott corrections".  Thus gradient
expansions must be generalized even for neutral atoms of large atomic number,
but these generalizations should recover the correct second-order limits.
Correlation is more subtle.
\end{abstract}


\maketitle


In a remarkable series of papers towards the end of his life,
Schwinger\cite{Sa80} 
(sometimes with Englert\cite{ESa84}) put the semiclassical theory
of neutral atoms on a firm footing.  
They carefully proved a variety
of results, including a clear-cut demonstration that the
local density approximation (LDA) becomes exact for exchange
as $Z$, the nuclear charge, tends to $\infty$.  
The large-$Z$ expansion of the energy of atoms whose rigor they established:
\ben
E = - 0.7687\, Z^{7/3} + 0.5\, Z^2 - 0.2699\, Z^{5/3} + \ldots
\label{EZasymp}
\een
is extremely close to the total (Hartree-Fock) energies of
neutral atoms (less than 0.5\% error for Ne), an example of
``the principle of unreasonable utility of asymptotic estimates."\cite{Sa80}
But their derivations are specific to neutral atoms,
and they eschew exploring any relation to ``the density-functional formalism,"\cite{ESa84}
preferring to express all quantities in terms of the potential.
Their results are little used within density functional theory (DFT).

In the quarter century since, 
Kohn-Sham DFT
has become a widely-used tool for electronic structure
calculations of
atoms, molecules, and solids\cite{FNM03}.
Here, the non-interacting kinetic energy, $T\s$, is treated exactly, and
only the density
functional for the exchange-correlation energy,
$E\xc[\n]$, must be approximated.
For $E\xc$, LDA\cite{HK64} is very reliable, but insufficiently
accurate for chemical bonding.  The gradient expansion, employing
density gradients, can be
derived from the slowly-varying electron gas for small gradients, but 
fails for real molecules and solids, whose gradients are not small.
The development of modern generalized gradient approximations (GGA)\cite{B88,PBE96},
depending on density gradients beyond that of leading order,
improved accuracy and led to the widespread use of DFT in many fields.
Anomalously, modern successful GGA's for exchange have gradient
coefficients that are about double that of the gradient expansion,
and the relevance of even LDA
to exponentially localized densities is often questioned\cite{PY89}.

But Thomas-Fermi (TF) theory\cite{T26,F28}, with its local approximation
to $T\s$, is the simplest, original
form of DFT, and yields the leading term in Eq. (\ref{EZasymp}).
Since $E=-T$ for atoms and correlation is O$(Z)$, Eq. (\ref{EZasymp})
is an expansion for $T\s$.
The Scott correction\cite{S52} ($Z^2$ term)
arises from the cusp at the nucleus and the $1s$-region electrons\cite{Sa80},
while the $Z^{5/3}$ term includes
second-order gradient contributions to $T_s$.

We introduce a methodology that generalizes
Schwinger's results to all systems and to other components of the energy.
It explains why local approximations become exact for large numbers
of electrons, and when the gradient expansion is accurate for real matter.
It explains the doubling of the coefficient for exchange, yielding a
nonempirical derivation of
the fit parameter of the B88 functional\cite{B88}, and
why the analog of Eq.(\ref{EZasymp}) fails for correlation.

To begin, define the scaled density
\ben
\n\z(\br) = \zeta^2\, \n(\zeta^{1/3}\br),~~~~~~~~0 < \zeta < \infty
\een
for any electron density.  If $\n(\br)$ contains $N$ electrons,
$\n\z(\br)$ contains $\zeta N$ electrons.   We only ever consider
integer $\zeta N$, but study energies as smooth functions of $\zeta$.
Changing $\zeta$ is exactly equivalent to changing $Z$ in TF theory,
and approximately so in reality.
This scaling is defined for all systems, not just atoms.

\begin{figure}[htb]
\unitlength1cm
\begin{picture}(12.5,5.5)
\put(-6,9){\makebox(12,6.5){
\includegraphics{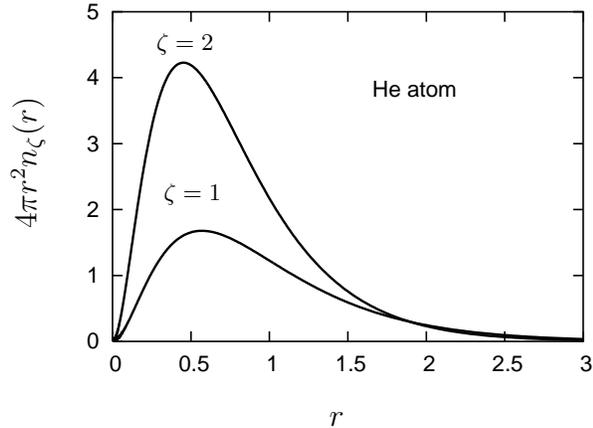}
}}
\setbox6=\hbox{\large $4\pi r^2 \n\z(r)$}
\put(0.8,3.5) {\makebox(0,0){\rotl 6}}
\put(4.8,0){\large $r$}
\put(2.6,3){$\zeta=1$}
\put(2.5,5){$\zeta=2$}
\end{picture}
\caption{Scaled radial density of He atom.}
\label{f:she}
\end{figure}
In Fig. \ref{f:she}, we illustrate the effect of the scaling on the density
of the He atom.  The density at $\zeta=2$ contains 4 electrons, has the
original shape, but is coordinate-scaled by $2^{1/3}$.
As $\zeta$ grows, the density becomes both large and slowly-varying
on the scale of the local Fermi wavelength, $\lambda_F(\br)=2\pi/k_F(\br)$, where 
$k_F(\br)=(3\pi^2\n(\br))^{1/3}$). 
Under this scaling, the dimensionless gradients $s=|\nabla \n|/(2 k_F \n)$
and $q=\nabla^2 n/(4 k_F^2 \n)$\cite{HK64,SGP82}
vary as
\ben
s\z(\br) ={s(\zeta^{1/3}\br)}/{\zeta^{1/3}},~~~~
q\z(\br) ={q(\zeta^{1/3}\br)}/{\zeta^{2/3}}.
\een
\begin{figure}[htb]
\unitlength1cm
\begin{picture}(12.5,5.5)
\put(-6,9){\makebox(12,6.5){
\includegraphics{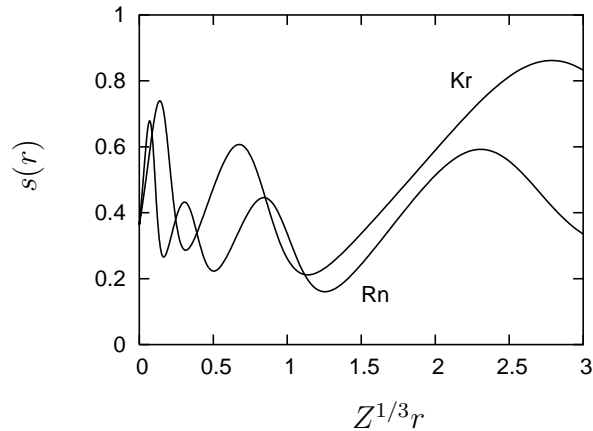}
}}
\setbox6=\hbox{\large $s(r)$}
\put(0.8,3.5) {\makebox(0,0){\rotl 6}}
\put(5.1,0){\large $Z^{1/3} r$}
\end{picture}
\caption{Reduced density gradient for noble gas atoms.}
\label{f:ssc}
\end{figure}
In Fig. \ref{f:ssc}, we plot $s$ 
for accurate densities (self-consistent OEP densities
with exact exchange) of both Kr and Rn. The gradient expansions
for the kinetic and exchange energies are expected \cite{HK64} to become exact as the
density $n(\vec{r})$ becomes slowly-varying on the scale of 
$\lambda_F$, i.e., when $s\to 0$ and $|q/s|\to 0$ everywhere.
The second condition is needed because $s\to 0$ for an infinitesimal-amplitude but
rapid variation around a uniform density.

Under $\zeta$-scaling of a mono- or
polyatomic density, as $\zeta\to\infty$ and as $s$ and $|q/s|\to 0$
(except in regions to be discussed),
the gradient expansion becomes asymptotically exact so that (aside
from possible oscillations\cite{ESa84}):
\bea
T\s[n\z]&=&\zeta^{7/3}\, T\s^{(0)}[n] + \zeta^{5/3}\, T\s^{(2)}[\n] 
+ \zeta\, T\s^{(4)}[\n]
+ \ldots,\nonumber\\
E\x[n\z]&=&\zeta^{5/3}\, E\x^{(0)}[\n]
+ \zeta\,  E\x^{(2)}[\n] + \ldots,
\label{asymp}
\eea
where $T\s^{(j)}[\n]$ is the $j$-th order contribution to the 
gradient expansion of $T\s$ (i.e., $T\s^{(0)}$ is the TF kinetic energy,
$T\s^{(2)}$ is 1/9 the von Weizs\"acker term\cite{Kc57}, etc.), and similarly 
for exchange\cite{AK85}.
The terms displayed in Eq. (\ref{asymp}) are those of the 
gradient expansion that remain finite for exponentially localized
densities, and also those
for which the gradient expansion
becomes asymptotically exact as $\zeta\to \infty$
for analytic densities (unless the density has nuclear cusps, when $T\s^{(4)}$ must
be excluded). The gradient expansions of
Eq. (\ref{asymp}) are ``statistical" approximations in the sense that they become
relatively exact as $\zeta N \to \infty$.

Evanescent regions are classically-forbidden regions in which the kinetic
energy density $\tau'(\br) = \sum_i \psi_i^*(\vec{r})^*\left( -\nabla^2/2\right)
\psi_i(\br)$
of the Kohn-Sham orbitals is negative and the gradient expansion must fail.
To investigate the contributions to $T\s[\n\z]$ and $E\x[\n\z]$ from evanescent
and nuclear cusp regions, we note that each can be represented by an 
exponential density $a\exp(-b r)$.  For this density, the evanescent region $r>r_e(\zeta)$
is defined loosely
by the conditions $s > 1$ and $|q/s| > 1$ (found from the second-order gradient expansion of
$\tau'(\vec{r})$).  Its radius $
r_e(\zeta)$ decreases like
$\ln(\zeta)/\zeta^{1/3}$ as $\zeta\to\infty$.  In this region, the 
considered terms of the gradient expansion 
each contribute of order $\zeta^{2/3}$ to $T\s$, and of order $\zeta^{1/3}$
to $E\x$ (with the convention that $\ln(\zeta)$ is of order 1).  
All of these contributions are of lower order than those
shown in Eq. (\ref{asymp}),
and thus asymptotically unimportant.  Similarly, the nuclear
cusp regions $r<r_c(\zeta)$ can be defined loosely by the condition $|q/s|>1$;  their 
radii $r_c(\zeta)$ decrease like $\zeta^{-2/3}$ as
$\zeta\to\infty$.  In these regions, the
local contributions dominate.   The local terms contribute
of order $\zeta^{4/3}$ to $T\s$ and $\zeta^{2/3}$ to $E\x$, and
are again asymptotically unimportant (once we truncate the gradient
expansions of Eq. (\ref{asymp})) at second order).
Our evanescent regions coincide with the edge surface of Ref. \cite{KMd98},
but the cusps do not.

  While the total electron number scales up as $\zeta$, the number in the
evanescent and cusp regions remains of order $\zeta^0$.  The {\em exact}
kinetic energy contributions of these regions are  of order $\zeta^{2/3}$ and
$\zeta$ (or $\zeta^{2/3}$, for the positive kinetic energy
density $\tau(\vec{r}) = \sum_i |\nabla \psi_i(\vec{r})|^2/2$), respectively.  All orders of 
$\zeta$
are unchanged if the bound 1 is replaced by a
smaller positive value, or if
the spherical evanescent region is replaced by a planar one.

To appreciate the significance of asymptotic exactness, 
$\zeta$-scale any density (even a single exponential).
Using modern linear response techniques,
construct the KS
potential and orbitals for each value of $\zeta$\cite{ZMP94}.
As $\zeta\to\infty$, deduce the local and gradient expansion
approximations for $T\s$ and $E\x$ exactly,
without ever studying the properties
of
the uniform or slowly-varying
electron gas.  
The system is becoming increasingly semi-classical, and the orbitals
well-described by their WKB approximations.  Semi-classical approximations
require only that the potential be smooth locally, not globally.

The behavior of correlation under this scaling is however recalcitrant.
In the bulk of the density, the gradient relative to the screening
length, $t=|\nabla n|/(2 k_s n)$, where 
$k_s(\br)={\sqrt{4k_F(\br)/\pi}}$, controls gradient corrections to correlation.
But
\ben
t\z(\br) = t(\zeta^{1/3}\br)
\een
does {\em not} change under our scaling, 
so that the density
does {\em not} become slowly-varying for correlation.
Nonetheless, the local approximation
still becomes exact, if the PBE GGA is a reliable guide (as motivated later): 
\ben
\label{corr}
E\c[\n\z] = A\c\, \zeta\, \ln\zeta + B\c\, \zeta + \ldots
\label{Echi}
\een
where $A\c =-0.02072$ is correctly given by LDA, while 
$B\c\LDA = -0.00452$ and  $B\c\PBE=0.03936$.
These numbers can be found by applying these
approximations to the TF density.
Because $t$ never gets small, $B\c$ has not only LDA and
GEA (gradient expansion to second order)
contributions, but higher-order contributions too.
This expansion is far more slowly converging than those of $T\s$ and $E\x$,
and much less relevant to real systems.

\begin{table}[htb]
\caption{Noble gas atomic XC energies compared with local and gradient
expansion approximations (hartree).}
\begin{tabular}{|c|ccc|ccc|}
\hline
&\multicolumn{3}{|c|}{$E\x$}
&\multicolumn{3}{|c|}{$E\c$}\\
\hline
 atom&  LDA&  GEA&  exact& LDA& GEA&exact\\
\hline
  He&   -0.884&   -1.007&      -1.026&   -0.113&    0.103&   -0.042\\
  Ne&      -11.03&    -11.77&  -12.10&   -0.746&    0.559&    -0.390\\
  Ar&    -27.86&    -29.29&    -30.17&   -1.431&     1.09&   -0.722\\
  Kr&    -88.62&    -91.65&    -93.83&   -3.284&     2.06&    --\\
  Xe&     -170.6&     -175.3&  -179.1&   -5.199&     3.15&    --\\
  Rn&     -373.0&     -380.8&  -387.4&   -9.026&     4.78&     --\\
\hline
\end{tabular}
\label{t:ExcLDA}
\end{table}

In Table I,
we list the exact\cite{CGDP93}, LDA, and GEA results for the noble
gas atoms using the known second-order gradient terms\cite{AK85,MB68}.
LDA gets relatively better and better as $Z$ grows,
and the GEA for exchange does even better, but the GEA for
correlation strongly overcorrects the LDA, making correlation energies
positive\cite{MB68}.
Thus this simple analysis explains why the gradient expansion yields
good answers for $T\s$ and $E\x$, but bad ones for 
correlation.  These explanations complement those already existing based
on holes and sum-rules\cite{PBW96}, but add the crucial ingredient that under this
scaling the gradient expansion becomes asymptotically exact for $T\s$ and $E\x$.

But Eq. (\ref{asymp}) for either $E\x$ or $T\s$ is far less accurate at $\zeta=1$ than
Eq. (\ref{EZasymp}) is for $E$ at $Z=1$.
Careful inspection of the origin in Fig. \ref{f:ssc} shows that the exact curves
approach a finite $s$-value at the origin, about 0.376, the hydrogenic
value.
The large-$Z$ expansion is not the same as scaling to large $\zeta$, except
in TF theory.  The gradients near the nucleus
do {\em not} become small on the Fermi wavelength scale, no matter how large
$Z$ is.  This region will contribute a term of order $Z^2$ to the kinetic
energy and of order $Z$ to the exchange energy at all levels, from LDA to exact.
LDA applied to the TF density produces
the leading term in $E\x$, $ 0.2208 Z^{5/3}$.  Gradient corrections are of order
$Z$, but so too is the cusp correction, i.e., the asymptotic expansion in
large $Z$ inextricably mixes these contributions (unlike in $T\s$).
Table I lists $E\x$ for noble gas atoms, 
calculated at the self-consistent non-relativistic OEP level.
We fit $(E_x+0.2208\, Z^{5/3})/Z$ as a
function of $Z^{-1/3}$, finding:
\ben
E\x(Z) = -0.2208\, Z^{5/3} + (C\LDA + \Delta C)\, Z + \ldots
\label{ExZ}
\een
Extraction of $C\LDA$ from LDA energies is difficult, because of shell-structure
oscillations.  We estimate $0 \gtrsim C\LDA \gtrsim -0.03$.
However, for any other calculation of $E\x$, we
find $(E\x-E\x\LDA)/Z$ is smooth, with fit results shown in Table II.
The large underestimate of GEA shows that gradient corrections
from the slowly-varying gas account for only half the entire contribution.
Assuming $C\LDA=0$,
$E\x(Z) \sim -0.2208\, Z^{5/3} -.196\, Z$.
This yields less than 10\% error for He, and less than 2\% for Ne.
\begin{table}[htb]
\caption{$\Delta C=\lim_{Z\to\infty} (E\x-E\x\LDA)/Z$ (hartree).
The ``gradient" contribution arises from the expansion to order
$\nabla^2$, while the remainder is the ``cusp".}
\begin{tabular}{|c|ccccc|}
\hline
$\Delta C$& GEA&  PBE&  B88&  TPSS& exact\\
\hline
total&-.098&-.174&-.202&-.159&-.196\\
\hline
gradient&-.098&-.174&-.217&-.0977&-.0977\\ 
cusp&0.00&-.000&+.015&-.0617&-.0979\\ 
\hline
\end{tabular}
\label{t:DC2}
\end{table} 

Popular GGA's such as PBE and B88\cite{B88} have second-order gradient coefficients
that are about twice the correct coefficient for a slowly-varying density, as
they must to reproduce accurate exchange energies of atoms.  But with
an incorrect coefficient, they cannot predict accurate
surface energies for metals\cite{CPT06}.
The origin of the enhanced gradient coefficient of the GGA for exchange 
now has a simple explanation.  In order to be asymptotically exact
for large $Z$, and hence accurate for most finite $Z$,
the functional accounts for both the slowly-varying term
and the cusp correction.
No GGA can get both effects right individually.
B88 is closest to being
exact for $\Delta C$, because of the fitting to
noble gas atoms\cite{B88}.
(In fact, assuming the exact gradient and cusp contributions
to $\Delta C$ are equal, asymptotic exactness
requires $\beta=5/(108 (6\pi^5)^{1/3})$ in B88, close to the fitted value of
.0042.)
PBE preserved the nearly correct uniform-gas linear response of LDA for XC together\cite{PBE96}, which
produces a $\Delta C$ also close to exact, and a GGA close to
that of a hole model\cite{PBW96}. A ``buried
1s" region has small s, looking to a GGA like a region of slowly-varying
density.  

However, a meta-GGA that employs $\tau(\br)$
can recognize that this is a rapidly-
varying region, and thus get everything right. The TPSS meta-GGA recovers the
gradient expansion to fourth order\cite{TPSS03}, while yielding a good
estimate for $\Delta C$ (Table II).

\begin{figure}
\unitlength1cm
\begin{picture}(12.5,5.5)
\put(-6,9){\makebox(12,6.5){
\includegraphics{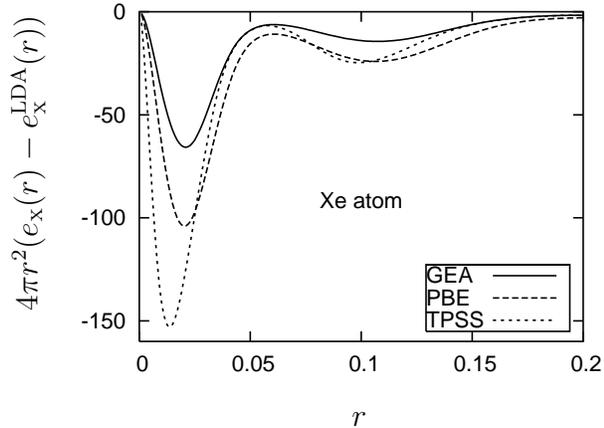}
}}
\setbox6=\hbox{\large $4\pi r^2\big(e\x(r)-e\x\LDA(r)\big)$}
\put(0.8,3.5) {\makebox(0,0){\rotl 6}}
\put(5.1,0){\large $r$}
\end{picture}
\caption{Radial exchange energy density differences for Xe
for different approximations (atomic units).}
\label{f:endens}
\end{figure}

Figure \ref{f:endens} plots the difference in 
exchange energy densities relative to LDA in the different
approximations.  PBE simply mimics GEA, being almost a factor
of 2 larger everywhere.  But TPSS produces a much greater contribution
from the region near the nucleus (via Fig. 1 of Ref. \cite{TPSS03}),
while reverting to the GEA value 
at the inner radii of the other atomic shells, where $s$ is small.

\begin{figure}[htb]
\unitlength1cm
\begin{picture}(12.5,5.5)
\put(-6,9){\makebox(12,6.5){
\includegraphics{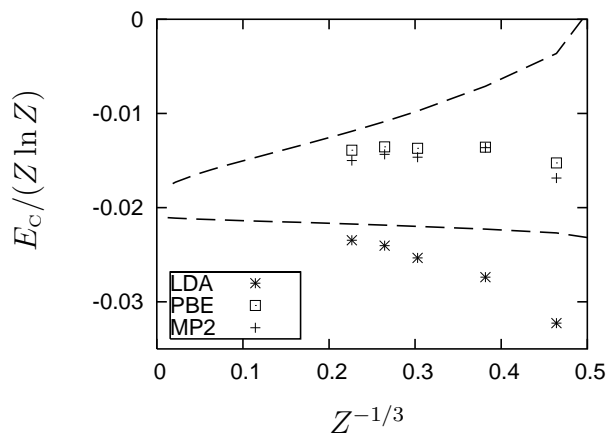}}}
\setbox6=\hbox{\large $E\c/(Z\ln Z)$}
\put(0.8,3.5) {\makebox(0,0){\rotl 6}}
\put(4.8,0){\large $Z^{-1/3}$}
\end{picture}
\caption{Scaled correlation energies for noble gas atoms (hartree).The dashed lines correspond to the
high-density limit of Eq. (\ref{corr}) for LDA and PBE.}
\label{f:Ecs}
\end{figure}
Again, correlation is less clear-cut.
In Fig. \ref{f:Ecs},
we plot the correlation energy
in three different approximations:
LDA, PBE, and M\/oller-Plesset second order perturbation theory
(MP2)\cite{FJS03}.  (TPSS is almost identical
on this scale to PBE.)  
Also included are dashed lines that correspond to the high-density
limit of Eq. (\ref{Echi}) using LDA and PBE inputs for $B\c$.
Real atoms are so far from the
asymptotic limit for correlation
that asymptotic exactness is much less relevant in this case.

\begin{table}[htb]
\caption{Noble gas atomic correlation energies (hartree).  The fixed-node Diffusion
Monte Carlo (DMC) values\cite{MDTN05} are upper bounds.}
\begin{tabular}{|c|cccccc|}
\hline
atom & LDA &  PBE&  TPSS& MP2& DMC & exact\\
\hline
He&-0.113& -0.042& -0.043& --&-0.042&-0.042\\
    Ne&         -0.743&     -0.351&      -0.354&     -0.388&      -0.376&-0.390\\
    Ar&         -1.424&     -0.707&      -0.711&     -0.709&      -0.667&-0.722\\
    Kr&         -3.269&     -1.767&      -1.771&     -1.890&	  -1.688&--\\
    Xe&         -5.177&     -2.918&      -2.920&     -3.089&	  -2.647&--\\
    Rn& -9.026&-5.325& -5.33 &            -5.745&&--\\
\hline
\end{tabular}
\label{t:C}
\end{table}

Table III shows the correlation energies of closed-shell atoms as
predicted by PBE, TPSS, and MP2
\cite{FJS03}, along with essentially 
exact values\cite{CGDP93}
where known.   The agreement among these values is generally
good.  The large-$Z$ limit is not problematic for PBE and TPSS, since 
$t(\br)$ remains bounded except near the cusp and in the tail.  

Last, we relate this scaling to others.  The most standard
is uniform coordinate scaling\cite{LP85} ($\gamma^3\, \n(\gamma\br)$), under
which $T\s/\gamma^2$ and $E\x/\gamma$ remain unchanged.
More recently, number-scaling, in which $\n(\br)$ becomes $\nu\, n(\br)$,
has been proposed\cite{CHb99}.
In the large $\nu$ limit, all gradients in the bulk become small on
both local length scales,
making even the gradient expansion for correlation
asymptotically exact.
The present scaling can be regarded as a product of these,
with $\nu=\gamma^3=\zeta$.  A slowly-varying
product with bounded density is $n(\nu^{-1/3} \mathbf{r})$, $\nu\rightarrow\infty$.

Our $\zeta$-scaling allows the results of Schwinger's derivations
to be applied throughout DFT, yielding insight into the performance of approximate
functionals.
Even for uncondensed matter,
such functionals should incorporate the second-order gradient expansions (although GGA
total exchange energies then degrade due to $1s$ regions).
Based on this, we have developed such a GGA for exchange which is currently being tested.
We thank Eberhard Engel for the use of his atomic OPMKS code, and NSF 
(CHE-0355405 and DMR-0501588) and the Norwegian Research Council (148960/432)
for support.

\end{document}